\documentstyle[12pt]{article}

\catcode`\@=11

\@addtoreset{equation}{section}
\def\theequation{\arabic{section}.\arabic{equation}}

\def\@normalsize{\@setsize\normalsize{15pt}\xiipt\@xiipt
\abovedisplayskip 14pt plus3pt minus3pt%
\belowdisplayskip \abovedisplayskip
\abovedisplayshortskip  \z@ plus3pt%
\belowdisplayshortskip  7pt plus3.5pt minus0pt}

\def\small{\@setsize\small{13.6pt}\xipt\@xipt
\abovedisplayskip 13pt plus3pt minus3pt%
\belowdisplayskip \abovedisplayskip
\abovedisplayshortskip  \z@ plus3pt%
\belowdisplayshortskip  7pt plus3.5pt minus0pt
\def\@listi{\parsep 4.5pt plus 2pt minus 1pt
            \itemsep \parsep
            \topsep 9pt plus 3pt minus 3pt}}

\def\underline#1{\relax\ifmmode\@@underline#1\else
        $\@@underline{\hbox{#1}}$\relax\fi}
\@twosidetrue

\relax

\catcode`@=12

\catcode`\@=11

\def\section{\@startsection{section}{1}{\z@}{3.5ex plus 1ex minus
   .2ex}{2.3ex plus .2ex}{\large\bf}}

\def\thesection{\Roman{section}.}

\def\appendix{\setcounter{section}{0}
        \def\thesection{APPENDIX }
        \def\theequation{\Alph{section}.\arabic{equation}}}

\def\FERMIPUB{}

\def\ps@headings{\def\@oddfoot{}\def\@evenfoot{}
\def\@oddhead{\hbox{}\hfill
        \makebox[.5\textwidth]{\raggedright\ignorespaces --\thepage{}--
        \hfill {\rm FERMILAB--Pub--\FERMIPUB}}}
\def\@evenhead{\@oddhead}
\def\subsectionmark##1{\markboth{##1}{}}
}

\catcode`\@=12

\relax

\def\figcap{\section*{Figure Captions\markboth
        {FIGURECAPTIONS}{FIGURECAPTIONS}}\list
        {Fig. \arabic{enumi}:\hfill}{\settowidth\labelwidth{Fig. 999:}
        \leftmargin\labelwidth
        \advance\leftmargin\labelsep\usecounter{enumi}}}
 \relax
\def\tablecap{\section*{Table Captions\markboth
        {TABLECAPTIONS}{TABLECAPTIONS}}\list
        {Table \arabic{enumi}:\hfill}{\settowidth\labelwidth{Table 999:}
        \leftmargin\labelwidth
        \advance\leftmargin\labelsep\usecounter{enumi}}}
 \relax
\def\reflist{\section*{References\markboth
        {REFLIST}{REFLIST}}\list
        {[\arabic{enumi}]\hfill}{\settowidth\labelwidth{[999]}
        \leftmargin\labelwidth
        \advance\leftmargin\labelsep\usecounter{enumi}}}
 \relax

\catcode`\@=11

\def\FERMIPUB{}

\def\ps@headings{\def\@oddfoot{}\def\@evenfoot{}
\def\@oddhead{\hbox{}\hfill
        \makebox[.5\textwidth]{\raggedright\ignorespaces --\thepage{}--
        \hfill {\rm FERMILAB--Pub--\FERMIPUB}}}
\def\@evenhead{\@oddhead}
\def\subsectionmark##1{\markboth{##1}{}}
}

\ps@headings

\relax
\newskip\humongous \humongous=0pt plus 1000pt minus 1000pt
\def\caja{\mathsurround=0pt}
\def\eqalign#1{\,\vcenter{\openup1\jot \caja
        \ialign{\strut \hfil$\displaystyle{##}$&$
        \displaystyle{{}##}$\hfil\crcr#1\crcr}}\,}
\newif\ifdtup

\def\beq{\begin{equation}}
\def\eeq{\end{equation}}

\def\beqn{\begin{eqnarray}}
\def\eeqn{\end{eqnarray}}
\relax

\def\G2{{\; \rm GeV/}c^2}
\def\G{\; \rm GeV}

\def\dotx{\dotx{\dot\overline{x}}}

\relax

\begin{document}
\hbadness=10000
\begin{titlepage}
\nopagebreak
\begin{flushright}

        {\normalsize KUCP-41 \\
        November,~1991}\\
\end{flushright}
\vfill
\begin{center}
\renewcommand{\thefootnote}{\fnsymbol{footnote}}
{\large \bf Quantum Conserved Charges and S-matrices \\
in N=2 Supersymmetric Sine-Gordon Theory}
\footnote{Presented by T.~Uematsu
at the YITP Workshop on Developments in Strings and Field Theories,
Kyoto, Japan, Sep.~9-12~(1991) ; to be submitted to
Prog.~Theor.~Phys.~Supplement}

\vfill

       {\bf Ken-ichiro Kobayashi}

       Institute for Nuclear Study, University of Tokyo, \\
       Tanashi, Tokyo 188, Japan \\
\vfill
       {and}
\vfill
       {\bf Tsuneo Uematsu}

       Department of Physics, College of Liberal Arts and Sciences, \\
       Kyoto University,~Kyoto 606,~Japan \\
\vfill

\end{center}

\vfill
\nopagebreak
\begin{abstract}
We study the quantum conserved charges and S-matrices of N=2 supersymmetric
sine-Gordon theory in the framework of perturbation theory formulated in
N=2 superspace. The quantum affine algebras ${\widehat {sl_{q}(2)}}$ and super
topological charges play important roles in determining the N=2 soliton
structure and S-matrices of soliton-(anti)soliton as well as soliton-breather
scattering.
\end{abstract}
\vfill
\end{titlepage}
\pagestyle{plain}
\newpage
\voffset = -2.5 cm

\leftline{\large \bf 1. Introduction}
\vspace{0.8 cm}

  Recently there has been increasing interest in massive integrable field
theories in two dimensions which can be regarded as perturbed systems of
conformal field theories.  Following the idea originally due to Zamolodchikov
[1], we apply a certain perturbation term $\lambda \int \Phi d^2 w$ where
$\lambda$ is the coupling constant, upon the action of the conformal field
theory as follows:
$$ S = S_{CFT} - \lambda \int d^2 w \Phi (w,{\bar w}) \eqno(1.1)$$

  In this perturbed system the conformal invariance no longer holds, but the
theory is still solvable as a massive theory, because of the existence of
infinitely many conservation laws.  The soliton theories like KdV, modified
KdV and sine-Gordon theories belong to such a class of solvable models.  They
possess factorizable S-matrices due to an infinite number of conservartion
laws.  Although these soliton theories are related with each other, here we
shall focus our attention on the sine-Gordon theory.

  Let us now recapitulate what have been known so far about sine-Gordon
theory as a perturbed conformal field theory.  First of all, we consider
the bosonic (N=0) sine-Gordon theory.  It has been realized that
the conformal minimal model perturbed
by $\Phi_{(1,3)}$ operator leads to the integrable restriction of sine-Gordon
theory [2-6].  In the Feigin-Fuchs representation, the $\Phi_{(1,3)}$
operator is given by $e^{-i\beta\phi}$ with
${1\over 2}\beta^2 = {m\over {m+1}}$ for the
central charge of the minimal unitary series : $c=1-{6\over {m(m+1)}}$, where
 $m=3,4,\cdots $.
Together with the screening operator $e^{i\beta\phi}$ the $\Phi_{(1,3)}$
forms the $\cos{\beta\phi}$ interaction term of sine-Gordon theory.

  Some time ago Sasaki and Yamanaka [7] studied the higher-spin conserved
charges which are polynomials in Virasoro operators at the classical levels
as well as at quantum levels. Eguchi and Yang derived the conserved charges
based on the perturbation theory $\grave{a}$ la Zamolodchikov. It was also
realized by LeClair, Smirnov and Eguchi-Yang that the the Hilbert space is
truncated for the rational values of the coupling constant [2-6]. It has
turned out that quantum group symmetry [8,9] plays an important role for
the truncation.

  Now it should be interesting to see whether incorporation of supersymmetry
will affect the solvability of sine-Gordon theory.  The N=1 supersymmetric
version of sine-Gordon theory can also be regarded as the N=1 superconformal
minimal model perturbed by $\Phi_{(1,3)}$ operator. The conservation laws and
S-matrices are discussed by Sasaki-Yamanaka [7], Mathieu [11],
Ahn-Bernard-LeClair [12] and Schoutens [13].

  Now the question is the following : What about the N=2 supersymmetric
extension of sine-Gordon theory?  As we will see, there exist somewhat
different features for N=2 case in contrast to N=0 and 1 cases.

  The classical equation of motion for the N=2 sine-Gordon theory is given
by the coupled equations [14]:
$$\eqalign{
& {\overline D}_{+}D_{+}\phi^{+} = g \sin \beta\phi^{-}  \cr
& {\overline D}_{-}D_{-}\phi^{-} = g \sin \beta\phi^{+}  \cr
}\eqno(1.2)$$
where $\phi^{+}$ and $\phi^{-}$ are the chiral and anti-chiral superfields
of N=2 supersymmtric theory.  $Z=(z,\theta^{+},\theta^{-})$
and ${\bar Z}=({\bar z},{\bar \theta}^{+},{\bar \theta}^{-})$ are holomorphic
and anti-holomorphic parts of the N=2 supercoordinates, respectively.  The
chiral and anti-chiral superfields satisfy the constraints
$D_{-} \phi^{+} = {\overline D}_{-}\phi^{+}=0$ and
$D_{+} \phi^{-} = {\overline D}_{+}\phi^{-}=0$, where  the N=2 supercovariant
derivatives are defined as $D_{\pm} = \partial/{\partial{\theta}^{\pm}} +
{1\over 2}\theta^{\mp}{\partial}_z$ and similar expressions for
${\overline D}_{+}$ and ${\overline D}_{-}$.
The chiral and anti-chiral superfields are composed of a complex
boson ${\varphi}^{\pm}$ and a complex fermion ${\psi}^{\pm}$,
${\bar \psi}^{\pm}$ and the auxiliary field $F^{\pm}$ as:
$$
\phi^{\pm}=\varphi^{\pm} + \theta^{\pm}\psi^{\mp} +
{\bar \theta}^{\pm}{\bar \psi}^{\mp} + \theta^{\pm}{\bar \theta}^{\pm}F^{\pm}
\eqno(1.3)$$
In terms of the component fields the equations of motion read
$$\eqalign{
& \partial_z\partial_{\bar z}\varphi^{+} = -g^2 \sin{\varphi^{+}}
\cos{\varphi^{-}}-g^2\psi^{+}{\bar \psi}^{+}\sin{\varphi^{-}} \cr
& \partial_{\bar z}\psi^{-} = g{\bar \psi}^{+}\cos{\varphi^{-}} \cr
& \partial_z{\bar \psi}^{-} = -g\psi^{+}\cos{\varphi^{-}} \cr
}\eqno(1.4)$$
where we set $\beta=1$ for simplicity.  In the basis of real components
${\varphi}^{\pm}={1\over \sqrt2}({\varphi}_1 \pm i{\varphi}_2)$
one of the bosonic parts of the equations of motion, for $\varphi_1$,
in the limit of vanishing fermion fields becomes a sine-Gordon equation and
the other one, for $\varphi_2$, becomes a hyperbolic sine-Gordon
(sinh-Gordon) equation.

The conservation laws were studied at the classical level [14,15], and
some lower-spin conserved charges were explicitly constructed as polynomials
in super-Virasoro generator(super energy-momentum tensor) in
Feigin-Fuchs-Miura form.

More systematic method to investigate classical conservation laws is
provided by Lie superalgebraic approach.  Recenlty Inami and Kanno studied
the N=2 super KdV and sine-Gordon theories in this framework [16].
They have shown that the N=2 supersymmetric sine-Gordon corresponds to the
$A(1,1)^{(1)}$ Lie superalgebra.

Now what about the conservation laws at the quantum level?
Does the factorization of S-matrix into bosonic and supersymmetric parts
[17,12,13] also hold in N=2 case ?
For this purpose we shall base our argument on the perturbation theory.
In this article, we shall also present the argument on exactness of first
order perturbation.

In the next section, we discuss conservation laws in the framework of
N=2 perturbation theory based on the superspace formalism.  In section 3,
we present the quantum conserved charges which generate the quantum group
symmetry denoted as $\widehat{sl_{q}(2)}$.  In section 4, we study the
super topological charges of N=2 supersymmetry and their relation with
the topological charge which belongs to the quantum group algebras.
This analysis is important for determining the N=2 soliton structure.
We present the S-matrices for soliton-(anti)soliton as well as
soliton-breather scattering in N=2 sine-Gordon theory in section 5.
The final section is devoted to conclusion and future problems.

\vspace{0.8 cm}
\leftline{\large \bf 2. N=2 Perturbation Theory and Conservation Laws}
\vspace{0.8 cm}

The action which leads to the classical equations of motion for the N=2
sine-Gordon theory is constructed by adding a chiral perturbation to the N=2
free action $S_{free}$ as
$$\eqalign{
& S = S_{free} - \lambda \int d^2 z \Phi ( z, {\bar z})  \quad , \cr
& \Phi ( z, {\bar z}) = \int  d^2 \theta^{+}( e^{i\beta \phi^{+}}
+ e^{-i\beta \phi^{+}}) + \int d^2 \theta^{-}( e^{i\beta \phi^{-}} +
e^{-i\beta \phi^{-}}) \cr
& = 2 \bigl ( \int d^2 \theta^{+} \cos{\beta\phi^{+}}
+ \int d^2 \theta^{-} \cos{\beta\phi^{-}} \bigr ) \cr
}\eqno(2.1)$$
where $\phi^{+}$ and $\phi^{-}$ are free chiral and anti-chiral superfields
satisfying the conditions: \hfill\break
${\overline D}_{+}D_{+}\phi^{+} = 0$ and ${\overline D}_{-}D_{-}\phi^{-} = 0$.
Hence they are decomposed into holomorphic and anti-holomophic
parts as $\phi^{\pm}=S^{\pm} + {\bar S}^{\pm}$.

  To the lowest order in perturbation theory, we get
$$\partial_{\bar z_1} A(Z_1, {\bar Z_1}) = \lambda \partial_{\bar z_1}
\bigl \{ \int d^2 z_2 d^2 \theta_2^{-} A(Z_1)
(e^{i\beta\phi^{-}(Z_2)} + e^{-i\beta\phi^{-}(Z_2)}) + ( - \rightarrow + )
\bigr \} \eqno(2.2) $$
Let us suppose that the operator product expansion of a operator $A$ and the
chiral perturbation term $\exp[i\beta S^{-}(Z_2)]$ is given by
$$A(Z_1)e^{i\beta S^{-}}(Z_2) \sim {\theta_{12}^{-}\over Z_{12}} \times
({\textstyle residue}) \eqno(2.3)$$
where$Z_{12}$ and $\theta_{12}^{\pm}$ stand for the invariant distances
in N=2 superspace
$$Z_{12} = z_1 - z_2 - {1\over 2}(\theta_1^{+}\theta_2^{-} +
\theta_1^{-}\theta_2^{+}) \quad, \quad \theta_{12}^{\pm} =
\theta_1^{\pm} - \theta_2^{\pm}
\eqno(2.4)$$
Now if the residue behaves as
$$ {\textstyle residue} \sim D_{+}X + D_{-}X'
\eqno(2.5)$$
for some operators $X$ and $X'$, then
the following charge
$$Q= \oint dz d\theta^{+}d\theta^{-} A(Z) \eqno(2.6)$$
turns out to be a conserved charge, provided that (2.6) is invariant under
the interchanges:
(i)$ + \leftrightarrow -$ and
(ii)$\beta \leftrightarrow -\beta$.

  Now we can classify conserved charges at the quantum level into two
categories. The first category includes regularized Virasoro polynomials
which form an infinite set of higher-spin conserved charges that assure
the integrability of the theory. The second one contains extra non-local
conserved charges which do not have classical analogues in contrast to the
first category and lead to the quantum group symmetry as will be discussed
in the later section.

  In ref.[14,15], we have presented some lower-spin conserved polynomial
charges in terms of the super stress tensor in Feigin-Fuchs-Miura form:
$$ T = :D_{+}S^{+}D_{-}S^{-}: - i \alpha ({\partial}S^{+} - {\partial}S^{-})
\eqno(2.7)$$
which consists of the N=2 superconformal generators: the U(1) charge, the
supercurrents and the energy-momentum tensor (Virasoro generator). In the
above equation we choose $\alpha = 1/\beta$ so that the vertex operator
$e^{i\beta S^{\pm}}$ should have the conformal weight ${1\over 2}$ as
the chiral screening operators. Then it turns out that the coefficients of
the Virasoro polynomials depend upon \lq\lq central charge \rq\rq given by
$c=3(1-2/\beta^2)$.

  Here one important observation is in order.  Although
we applied the N=2 superconformal field theory (SCFT) technique, our N=2
sine-Gordon theory should not be regarded as a perturbed N=2 SCFT minimal
models, in contrast to the N=0 and N=1 cases.  This is because, if so,
a part of the chiral perturbation term $e^{-i\beta S^{\pm}}$ would possess
a negative conformal weight $-1/2$ which, of course, cannot be accepted.  We
should rather interpret the present N=2 sine-Gordon theory as a
super-renormalizable theory with zero background charge, for which the
coupling constant $\lambda$ has a mass dimension.

\vspace{0.8 cm}
\leftline{\large \bf 3. Quantum Conserved Charges}
\vspace{0.8 cm}

  As we have mentioned, the quantum theory has extra non-local charges of the
vertex operator type in addition to the polynomial charges. The vertex type
charges have no classical analogues.  In our N=2 sine-Gordon we found the
following extra quantum charges:
$$Q^{\pm}(\beta) = \oint dz d \theta^{+} d \theta^{-} :e^{\pm i{1\over \beta}
( S^{+} + S^{-} )}: = \oint dz J^{\pm}(z)
\eqno(3.1)$$
which can be seen to satisfy the condition (2.5).  As for $Q^{-}$, for
example, we find
$$\eqalign{
&:\exp \bigl [ -{i\over \beta}( S^{+} + S^{-}) \bigr ] (Z_1):
:\exp (i\beta S^{-})(Z_2): \cr
& \sim {\theta_{12}^{-}\over Z_{12}} \times
D_{-} \bigl \{ {1\over {1-\beta^2}}\exp [ -{i\over \beta}S^{+} +
i(-{1\over \beta}+\beta)S^{-} ](Z_2) \bigr \} \cr
}\eqno(3.2)$$
In eq.(3.1) the non-local conserved currents $J^{\pm}(z)$ are found
to be
$$J^{\pm}(z) =  \bigl ( \pm {1\over {\sqrt{2}\beta}} \partial_z
{\varphi}_2 + {1\over {\beta}^2}\psi^{+} \psi^{-} \bigr )
e^{\pm{i\sqrt{2}\over \beta} \varphi_1}
\eqno(3.3)$$
and similarly we have the anti-holomorphic non-local currents
$${\bar J}^{\pm}({\bar z}) = \bigl ( \mp {
1\over {\sqrt{2}\beta}} \partial_{\bar z} {\bar \varphi}_2
+ {1\over {\beta}^2}{\bar \psi}^{+} {\bar \psi}^{-} \bigr )
e^{\mp{i\sqrt{2}\over \beta} {\bar \varphi}_1}
\eqno(3.4)$$
Under the perturbation term $\lambda\int d^2 w \Phi(w,{\bar w})$
where
$$\Phi(w,{\bar w})=\sum_{a,b=+,-}\Phi^{(ab)}(w){\bar \Phi}^{(ab)}({\bar w})
\eqno(3.5)$$
with
$$\eqalign{
&\Phi^{(+ \pm)}(w)= \pm i\beta\psi^{+}(w)
e^{\pm i\beta\varphi^{-}}(w) \quad , \quad
\Phi^{(- \pm)}(w)= \pm i\beta\psi^{-}(w)
e^{\pm i\beta\varphi^{+}}(w) \cr
&{\bar \Phi}^{(+ \pm)}({\bar w})= \pm i\beta{\bar \psi}^{+}({\bar w})
e^{\pm i\beta{\bar \varphi}^{-}}({\bar w}) \quad , \quad
{\bar \Phi}^{(- \pm)}({\bar w})= \pm i\beta{\bar \psi}^{-}({\bar w})
e^{\pm i\beta{\bar \varphi}^{+}}({\bar w}) \cr
}\eqno(3.6)$$
they satisfy the following conservation laws
$$\partial_{\bar z}J^{\pm} = \partial_z H^{\pm} \quad ,
\quad \partial_z {\bar J}^{\pm} = \partial_{\bar z}{\bar H}^{\pm}
\eqno(3.7)$$
where
$$\eqalign{
&H^{\pm} = \lambda \sum_{ab}h^{\pm (ab)}(z)
{\bar \Phi}^{(ab)}({\bar z}) \cr
&{\bar H}^{\pm} = \lambda \sum_{ab}\Phi^{(ab)}(z)
{\bar h}^{\pm (ab)}({\bar z}) \cr
}\eqno(3.8)$$
together with the operator product expansion
$$\eqalign{
&J^{\pm}(z)\Phi^{(ab)}(w) \sim {1\over {z-w}}\partial_w h^{\pm (ab)}(w) \cr
&{\bar J}^{\pm}({\bar z}){\bar \Phi}^{(ab)}({\bar w}) \sim
{1\over {{\bar z}-{\bar w}}}\partial_{\bar w}
{\bar h}^{\pm (ab)}({\bar w}) \cr
}\eqno(3.9)$$
and the non-vanishing $h^{\pm(ab)}(w)$'s
are found to be
$$\eqalign{
& h^{+(\pm -)}(w)={{\pm i}\over \beta}\psi^{\pm}(w)e^{i({\sqrt{2}\over \beta}
-{\beta\over \sqrt{2}})\varphi_1 \mp {\beta\over \sqrt{2}}\varphi_2}(w) \cr
& h^{-(\pm +)}(w)={{\pm i}\over \beta}\psi^{\pm}(w)e^{-i(
{\sqrt{2}\over \beta}
-{\beta\over \sqrt{2}})\varphi_1 \pm {\beta\over \sqrt{2}}\varphi_2}(w) \cr
}\eqno(3.10)$$
and we also have similar expressions for ${\bar h}^{\pm(ab)}({\bar w})$'s.
{}From the conservation laws (3.7) we redefine conserved charges as
$$\eqalign{
& Q_{\pm} = \int dz J^{\pm} + \int d{\bar z} H^{\pm} \cr
& {\bar Q}_{\pm} = \int d{\bar z} {\bar J}^{\pm} + \int dz {\bar H}^{\pm} \cr
}\eqno(3.11)$$
It turns out that these quantum conserved charges generate the q-deformation
of $sl(2)$ affine Kac-Moody algebra denoted as $\widehat{sl_{q}(2)}$ which
has been discussed by Bernard and LeClair [10] for the N=0 sine-Gordon case.

  For the commutation relation between $Q_{+}$ and $Q_{-}$, for example, we
find by using the perturbation theory
$$Q_{+}{\bar Q}_{-} -q^{-2}{\bar Q}_{-}Q_{+}
= \lambda \int_{t} dx \partial_x K
\eqno(3.12)$$
where the quantum group deformation parameter $q$ is given by
$$ q = -e^{-{i \pi}/\gamma}
\quad , \quad \gamma = \beta^2
\eqno(3.13)$$
and the $1/\gamma$ is equal to the spin of the conserved charges $Q_{\pm}$.
Let us note that the $\gamma$ and $\beta$ are related with each other for
N=0 and N=1 cases as
$$
\gamma = {\beta^2 \over {2-\beta^2}} \quad (N=0)\quad , \quad
\gamma = {2\beta^2 \over {1-\beta^2}} \quad (N=1)
\eqno(3.14)$$
The quantity $K$ on the right-hand side of (3.12) is found to be
$$\eqalign{
\qquad K = & \bigl (-{1\over \beta^2} \bigr )\bigl \{
\exp  [ i ({{\sqrt{2}}\over {\beta}} - {{\beta}\over {\sqrt{2}}})
\phi_1(x,t) - {{\beta}\over {\sqrt{2}}}\phi_2(x,t)+i\phi_3(x,t)] \cr
& \qquad + \exp [ i ({{\sqrt{2}}\over {\beta}} - {{\beta}\over {\sqrt{2}}})
\phi_1(x,t) + {{\beta}\over {\sqrt{2}}}\phi_2(x,t)-i\phi_3(x,t)] \bigr \} \cr
}\eqno(3.15)$$
where we have bosonized the fermions as $\psi^{\pm}=e^{\pm i \varphi_3}$ and
${\bar \psi}^{\pm}=e^{\pm i {\bar \varphi}_3}$ in order to avoid the subtlety
of the Grassmann variables.
In eq.(3.15) we denote
$\phi_{i}(x,t)=\varphi_{i}(z) + {\bar \varphi}_{i}({\bar z})$ for i=1,2 and 3.

\vspace{0.8 cm}
\leftline{\large \bf 4. Topological Charges and Quantum Group Symmetry}
\vspace{0.8 cm}

  In this section, we shall show that the quantum conserved charges we found
in the previous section and the topological charge we now discuss generate a
quantum group symmetry.

  First we note that the action (2.1) is invariant under the following shifts
of the fields ($m$ being an integer):
\begin{description}
\item{1)}
$\phi_1 \rightarrow \phi_1 + {{2m\pi\sqrt2}\over \beta}$, $\quad$ $\phi_2$,
 $\phi_3$: fixed,
\item{2)}
$\phi_1 \rightarrow \phi_1 + {{(2m+1)\pi\sqrt2}\over \beta}$, $\quad$
$\phi_2$: fixed,
$\quad$ $\phi_3 \rightarrow \phi_3 + \pi$ (mod $2\pi$).
\end{description}

Then we find that there exists a topological current for our N=2 case :
$${\cal T}^{\mu}(x,t) = {\beta\over 2\pi} \epsilon^{\mu\nu}
\partial_{\nu} \{ {\textstyle \sqrt2} \phi_1(x,t) \}
\eqno(4.1)$$
where $\epsilon^{\mu\nu}$ is the Levi-Civita symbol in two dimensions.

Therefore we can define the toplogical charge as
$${\cal T} = \int_{-\infty}^{+\infty} {\cal T}^{0}dx
= {\beta\over 2\pi}{\textstyle \sqrt2} \{\phi_1(x=+\infty) -
\phi_1(x=-\infty)\}
\eqno(4.2)$$
The right-hand side of (3.12) turns out to be
$$\eqalign{
& \lambda \int_{t} dx \partial_x K \cr
& = 2\lambda/(-\beta^2) \bigl [
1 - e^{i({\sqrt2\over \beta}-
{\beta\over \sqrt2})
\phi_1(x=-\infty)} \cos \phi_3(x=-\infty) \bigr ] \cr
}\eqno(4.3)$$
where we have taken the soliton configurations
$\phi_1(x=+\infty)=\phi_3(x=+\infty)=0$ which derives from the translational
invariance, and $\phi_2(x=+\infty)=\phi_2(x=-\infty)=0$ because the hyperbolic
sine-Gordon component damps at $\pm\infty$.  Therefore, the topological
charge is determined by $\phi_1(x=-\infty)$ as
$$ {\cal T}=-{\beta\over 2\pi}{\textstyle \sqrt2}\phi_1(x=-\infty)
\eqno(4.4)$$
Hence we have the following boundary conditions in accordance with the
invariance of the action:

  i) ${\cal T}$ is an odd integer, $\phi_3(x=-\infty)=\pi$
(mod $2\pi$)  and

ii) ${\cal T}$ is an even integer, $\phi_3(x=-\infty)=0$
(mod $2\pi$). \hfill\break
Therefore we find the quantum commutation relation as follows
$$\eqalign{
&Q_{\pm}{\bar Q}_{\pm} -q^{2}{\bar Q}_{\pm}Q_{\pm} = 0
\cr
&Q_{\pm}{\bar Q}_{\mp} -q^{-2}{\bar Q}_{\mp}Q_{\pm} =a(1-q^{\pm 2{\cal T}})
\cr
[{\cal T}, Q_{\pm} ] &= \pm 2Q_{\pm} \quad , \quad
[{\cal T}, {\bar Q}_{\pm} ] = \pm 2{\bar Q}_{\pm}
\cr
}\eqno(4.5)$$
where we have defined a constant $a=2\lambda/(-\beta^2)$.
{}From this commutation relation we can easily see that the extra charges
generate the quantum affine algebra ${\widehat {sl_{q}(2)}}$ with vanishing
center, which has been discussed by Bernard and LeClair [10].

  In the above derivation of (4.5), we have computed the algebras of the
non-local quantum conserved charges to the first order in perturbation theory.
Here one should note that the first order correction is the only possible
correction for general values of $\beta$ and hence the above result is
actually exact. This statement was not explicitly shown in the previous
articles [14,15,22].
This can be shown by writing down all the possible
perturbation terms and considering their scale dimensions as well as their
symmetry. The proof goes as follows.  Let us take the case of
$\partial_{\bar z} J^{\pm}$, for illustration. We find that the
left(holomorphic)
and right(anti-holomorphic) conformal dimensions of
$\partial_{\bar z} J^{\pm}$
is given by $(1+{1\over \beta^2},1)$.  Then the possible $\lambda^n$ (n-th
order) term should have the form
$$\psi{\bar \psi}e^{{i\over \beta}\varphi^{+}+{i\over \beta}\varphi^{-}+
il\beta\varphi^{+}+im\beta\varphi^{-}}
e^{il\beta{\bar \varphi}^{+}+im\beta{\bar \varphi}^{-}}
\eqno(4.6)$$
with $k$ $\partial_z$'s and $k'$ $\partial_{\bar z}$'s being multiplied in
front. In the above equation $\psi$ (${\bar \psi}$) denotes either
$\psi^{+}$ (${\bar \psi}
^{+}$) or $\psi^{-}$ (${\bar \psi}^{-}$).
Terms without fermion fields would not lead to conserved charges that are
invariant under supersymmetry transformations. By noting that the scale
dimension of $\lambda$ is $({1\over 2},{1\over 2})$, we have the following
equations
$$\eqalign{
&  {1\over 2} + lm\beta^2 + {1\over \beta^2} + l + m +{n\over 2}
+ k = 1 + {1\over \beta^2} \cr
& {1\over 2} + lm\beta^2 + {n\over 2} + k' = 1 \cr
}\eqno(4.7)$$
{}From the above equations we get $lm=k'=0$ and $n=1$.  Therefore only the
first
order term is allowed from the dimensional counting.  This means that the
lowest order term gives in fact the exact result.  This kind of argument also
applies to more complicated cases like (4.5).

  We now consider the topological charges of N=2 supersymmetry which are
different from the ordinary sine-Gordon topological charge given in (4.2).
The supercharges are holomorphic without the perturbation. But now we should
consider the effect of the perturbation terms.  In the same methods as the
bosonic topological charge, it is easily checked that the superalgebra
has the topological modification [18].

The holomorphic and anti-holomorphic supercurrents are written in terms of
components;
$$\eqalign{
& G^{\pm} (z)=\psi^{\pm} (z) \partial_z \varphi^{\pm} (z) \cr
& {\bar G}^{\pm} ({\bar z})={\bar \psi}^{\mp} ({\bar z}) \partial_{\bar z}
{\bar \varphi}^{\mp} ({\bar z}) \cr
}\eqno(4.8)$$
which obey the conservation laws in perturbation theory as in the extra
non-local currents:
$$\partial_{\bar z}G^{\pm} = \partial_z F^{\pm} \quad ,
\quad \partial_z {\bar G}^{\pm} = \partial_{\bar z}{\bar F}^{\pm}
\eqno(4.9)$$
where
$$\eqalign{
&F^{\pm} = \lambda \sum_{ab}f^{\pm(ab)}(z){\bar \Phi}^{(ab)}({\bar z}) \cr
&{\bar F}^{\pm} = \lambda \sum_{ab} \Phi^{(ab)}(z){\bar f}^{\pm(ab)}
({\bar z})
\cr
}\eqno(4.10)$$
with $f^{\pm(ab)}(z)$'s and ${\bar f}^{\pm(ab)}({\bar z})$'s being defined
through the operator product expansion:
$$\eqalign{
& G^{\pm}(z)\Phi^{(ab)}(w) \sim {1\over {z-w}}\partial_w f^{\pm(ab)}(w) \cr
& {\bar G}^{\pm}({\bar z}){\bar \Phi}^{(ab)}({\bar w}) \sim
{1\over {{\bar z}-{\bar w}}}\partial_{\bar w} {\bar f}^{\pm(ab)}({\bar w}) \cr
}\eqno(4.11)$$
Now we define the supercharges as follows
$${\cal Q}^{\pm}=\int dz G^{\pm} + \int d{\bar z} F^{\pm} \quad , \quad
{\bar {\cal Q}}^{\pm}=\int d{\bar z} {\bar G}^{\pm}
+ \int dz {\bar F}^{\pm}
\eqno(4.12)$$
and non-vanishing  $f^{\pm(ab)}$'s and ${\bar f}^{\pm(ab)}$'s are given by
$$\eqalign{
& f^{+(-\pm)}=-e^{\pm i\beta\varphi^{+}} \quad , \quad
f^{-(+\pm)}=-e^{\pm i\beta\varphi^{-}} \cr
& {\bar f}^{+(+\pm)}=-e^{\pm i\beta{\bar \varphi}^{-}} \quad , \quad
{\bar f}^{-(-\pm)}=-e^{\pm i\beta{\bar \varphi}^{+}} \cr
}\eqno(4.13)$$
Perturbation theory leads to the following anti-commutation relation
$$
\bigl \{ {\cal Q}^{\pm}, {\bar {\cal Q}}^{\mp} \bigr \}
= \lambda \int dx \partial_x K^{\pm\mp}\eqno(4.14)$$
where
$$\eqalign{
K^{\pm\mp} & = \sum_{ab}f^{\pm(ab)}(z){\bar f}^{\mp(ab)}({\bar z})
= e^{i\beta(\varphi^{\pm} + {\bar \varphi}^{\pm})}
+ e^{-i\beta(\varphi^{\pm} + {\bar \varphi}^{\pm})} \cr
& = \bigl [ e^{i{\beta\over \sqrt{2}}(\phi_1(x,t) \pm i\phi_2
(x,t) )} +
e^{-i{\beta\over \sqrt{2}}(\phi_1(x,t) \pm i\phi_2
(x,t) )} \bigr ] \cr
}\eqno(4.15)$$
while we get
$$\bigl \{ {\cal Q}^{\pm}, {\bar {\cal Q}}^{\pm} \bigr \}=0 \eqno(4.16)$$
With the same soliton configuration as discussed before we obtain
$$\bigl \{ {\cal Q}^{\pm}, {\bar {\cal Q}}^{\mp} \bigr \}
= \lambda \bigl [ 2 - ( e^{i{\beta\over \sqrt2}\phi_1(-\infty)}
+ e^{-i{\beta\over \sqrt2}\phi_1(-\infty)}) \bigr ] = {\cal T}' \eqno(4.17)$$
Hence we note that the super topological charge ${\cal T}'$ is determined
by the topological charge ${\cal T}$ as
$$ {\cal T}' = 2\lambda ( 1 - (-1)^{\cal T})\eqno(4.18)$$
Therefore, when ${\cal T}$ is an odd integer we get ${\cal T}'=4\lambda$,
 on the other hand, if ${\cal T}$ is an even integer we have ${\cal T}'=0$.
Since the soliton and the anti-soliton possess the topological charge
${\cal T}= \pm 1$, they must have non-zero super topological charge
${\cal T}'=4\lambda$.  We will see the physical consequence of this fact in
the next section.

In terms of the real basis $Q_1$ and $Q_2$ satisfying
$$\{Q_{1,2}, \bar{Q}_{1,2}\}=2T_{1,2} \eqno(4.19)$$
${\cal Q}^{\pm}$ and ${\bar {\cal Q}}^{\pm}$ are given by
$${\cal Q}^{\pm}= \textstyle{1\over \sqrt2}(Q_1 \pm i Q_2) \quad , \quad
 {\bar {\cal Q}}^{\pm}= \textstyle{1\over \sqrt2}({\bar Q}_1 \mp i {\bar Q}_2)
\eqno(4.20)$$
and therefore we have ${\cal T}'=T_1 - T_2 =2T_1 = -2T_2$.

\vspace{0.8 cm}
\leftline{\large \bf 5. Soliton Structure and S-matrix of N=2 Sine-Gordon
Theory}
\vspace{0.8 cm}

  In the present section, we examine the N=2 soliton structure and construct
the S-matrices of the N=2 sine-Gordon theory.

  We first study the realization of N=2 supersymmetry.  Let us first consider
the realization of N=1 supersymmetry as discussed by Zamolodchikov [19].
The tri-critical Ising model possesses N=1 supersymmetry and is described by
$\phi^{6}$ potential in the Landau-Ginzburg picture.  The suitably perturbed
model which is still solvable and has N=1 supersymmetry is represented as
the deformed $\phi^{6}$ potential which has three minima.  We denote these
minima by numbers, -1, 0, 1.  The fandamental particles correspond to the
soliton or kink states which are classical configurations going from one
minimum to another.  We denote a kink state connecting the degenerate vacua
$a$ and $b$ ($a,b=0,\pm 1$, $|a-b|=1$) by $|K_{ab}(\theta)\rangle$ where
$\theta$ represents the rapidity of the state.
Therefore we have four solitons which we denote by $|K_{1 0}\rangle$,
$|K_{-1 0}\rangle$,
$|K_{0 1}\rangle$ and $|K_{0 -1}\rangle$.
We now introduce supercharges denoted as $Q$ and $\bar{Q}$. The supercharge
$Q$ acts on this kink state as
$$Q|K_{ab}(\theta)\rangle =i(a+ib)e^{\theta/2}K_{-ab}(\theta)
\eqno(5.1)$$
where $a^2+b^2=1$, and we have similar expression for ${\bar Q}$ by replacing
$i(a+ib)$ with $-i(a-ib)$. We easily see that the superalgebra is realized as:
$$Q^2=e^{\theta}=P, \quad \bar{Q}^2=e^{-\theta}=\bar{P},
\quad \{Q,\bar{Q}\}=2T \eqno(5.2)$$
where we have taken the mass to unity and $T$ is
the super-topological charge.
Any multi-particle state can be given by
$$|K_{a b}(\theta_{1})K_{b c}(\theta_{2})K_{c d}(\theta_{3}) \cdots \rangle
\eqno(5.3)$$
The supersymmetry is realized on the multi-particle state as
$$\eqalign{
& Q | K_{a_{1} a_{2}}(\theta_{1}) K_{a_{2} a_{3}}(\theta_{2}) \cdots
  K_{a_{N-1} a_{N}}(\theta_{N-1})\rangle = \cr
& \sum_{i=1}^{N}  i(a_{i}+ia_{i+1}) e^{\theta_i/2}
|K_{-a_1 -a_2}\cdots K_{-a_{i-1} -a_{i}}
K_{-a_{i} a_{i+1}} K_{a_{i+1} a_{i+2}} \cdots K_{a_{N-1} a_{N}}\rangle \cr
}\eqno(5.4)$$

 Now we proceed to the N=2 supersymmetry which can be realized as a tensor
product of two tri-critical Ising soliton states.  The fundamental particles
are written in the form:
$$|K_{a b, c d}\rangle \equiv |K_{a b} K_{c d}\rangle \eqno(5.5)$$
  In N=2 case, we have supercharges $Q_1$ and $Q_2$ and conjugated ones.  We
assume that the first supercharge operates on the first factor of the right
hand side of the equation just as tri-critical Ising model.  But when the
second supercharge operates on the second factor, we require that the sign of
the components of the first factor should always change.  For example, the
supercharge acts on the one-particle states:
$$\eqalign{
& Q_{1}|K_{ab, cd}\rangle=i(a+ib)e^{\theta/2}|K_{-ab, cd}\rangle, \quad
  |\bar{Q_{1}}K_{ab, cd}\rangle=-i(a-ib)e^{-\theta/2}|K_{-ab, cd}\rangle, \cr
& Q_{2}|K_{ab, cd}\rangle=i(c+id)e^{\theta/2}|K_{-a -b, -cd}\rangle, \quad
  |\bar{Q}_{2}K_{ab, cd}\rangle=-i(c-id)e^{-\theta/2}|K_{-a -b, -cd}\rangle
\cr
}\eqno(5.6)$$
Therefore the supercharges satisfy
$$\eqalign{
& Q_{1}^{2}=Q_{2}^{2}=e^{\theta}=P,\quad \bar{Q}_{1}^{2}=\bar{Q}_{2}^{2}
=e^{-\theta}=\bar{P},
 \cr
& \{Q_{1,2}, \bar{Q}_{1,2}\}=2T_{1,2}, \quad T_1=-(a^2 - b^2), \quad
T_2=-(c^2 - d^2) \cr
} \eqno(5.7)$$
and they, otherwise, anticommute with each other.  The reason why we
demand that the sign in the first factor change when we operate the second
charge is to assure the anticommutativity of $Q_{1}$ and $Q_{2}$.

  Now the bosonic and fermionic N=1 tri-critical Ising soliton states and
their anti-particle states are constructed as [12,13,20]:
$$\eqalign{
&|B\rangle = {1\over {\textstyle \sqrt2}}(|K_{-10}\rangle + |K_{10}\rangle )
\quad , \quad
|F\rangle = {1\over {\textstyle \sqrt2}}(|K_{-10}\rangle - |K_{10}\rangle )
\cr
&|{\bar B}\rangle = {1\over {\textstyle \sqrt2}}
(|K_{0-1}\rangle + |K_{01}\rangle )
\quad , \quad
|{\bar F}\rangle = {1\over {\textstyle \sqrt2}}
(|K_{0-1}\rangle - |K_{01}\rangle )
\cr
}\eqno(5.8)$$

  As we mentioned, our quantum conserved charges are integrals of the highest
component of their supermultiplet by construction. Therefore they commute
with the supercharges.  This means that we have two commuting symmetries,
i.e. N=2 supersymmetry and quantum affine algebra ${\widehat {sl_{q}(2)}}$.
Hence the S-matrix can be written as a product of \lq\lq minimal
supersymmetric S-matrix \rq\rq and the ordinary sine-Gordon S-matrix
as in the N=1 case discussed by Ahn et al.[12,20] and by Schoutens [13].
$$S_{N2SG} \sim S_{MN2} \times S_{SG} \eqno(5.9)$$
where $S_{N2SG}$, $S_{MN2}$ and $S_{SG}$ represent the S-matrices of N=2
supersymmetric sine-Gordon, of the minimal N=2 supersymmetric model,
and of ordinary bosonic sine-Gordon [21], respectively.
We expect the \lq\lq minimal supersymmetric S-matrix \rq\rq is that of
the coset model $SU(2)_{2} \times SU(2)_{2} / SU(2)_{4}$
which is the perturbed c=1 model discussed in [12,20].  Alternatively, it
may be realized at a special value of the coupling constant,
$\beta={2\over \sqrt{3}}$, of the bosonic sine-Gordon theory
which corresponds to c=1 N=2 super CFT [15].

  The fundamental particles of our model can be represented as the product
of RSOS solitons of the coset model and kinks of the sine-Gordon part.  We
can show that they acctually form the super multiplets, though we need
suitable truncation to obtain the minimal number of particles.

  Fundamental particles that form supermultiplets are
$$ |BB\rangle, \quad |BF\rangle, \quad |FB\rangle, \quad |FF\rangle,
\quad |B\bar{B}\rangle,
\quad |B\bar{F}\rangle, \quad |F\bar{B}\rangle, \quad |F\bar{F}\rangle
\eqno(5.10)$$
and their conjugates.  So we have sixteen fundamental particles.  But some
consideration about topological charges leads to reduction of the number of
fundamental particles.  Because of the form of the potential without fermions,
we expect kink and anti-kink solutions as in the case of bosonic and N=1
supersymmetric cases.  And the topological charges are discussed in [12,13]
and in the beginning of this paper.  Since kink and anti-kink solutions have
odd topological charge, their super topological charges should be non-zero.
This restricts $|XY\rangle$ in eq.(5.10) to one of $$|B{\bar B}\rangle, \quad
|B{\bar F}\rangle, \quad |F{\bar B}\rangle, \quad |F{\bar F}\rangle
\eqno(5.11)$$
which possess $T_1=-T_2=-1$ and hence ${\cal T}'=-2$ and their conjugates
with
$T_1=-T_2=1$ and ${\cal T}'=2$.
For remaining states the super-topological charges vanish.
In fact, particles of (5.11) and their conjugates form a closed set, i.e.
we can restrict on-shell states to the set and other states cannot appear
in the final state.  Of course, this is another expression of the conservation
laws of super topological charges.   Therefore we have four supermultiplets;
those of
$$|A^{\pm}B{\bar B}\rangle, \quad |A^{\pm}\bar{B}B\rangle
\eqno(5.12)$$
where we have denoted kink and anti-kink states of ordinary sine-Gordon
 theory by $A^{+}$ and $A^{-}$ [21].

  Now we parametrize the S-matrices for tri-critical Ising solitons
scattering : \hfill\break $K_{ab}(\theta_1) + K_{bc}(\theta_2) \rightarrow
K_{ad}(\theta_2) + K_{dc}(\theta_1)$ as
$$|K_{a b}(\theta_{1})K_{b c}(\theta_{2})\rangle_{in} =
\sum_d  S^{da}_{bc} (\theta_{12})|K_{a d}(\theta_{2}) K_{d c}(\theta_{1})
\rangle_{out}
 \eqno(5.13)$$
where $a,b,c,d = 0, \pm1$, and $\theta_{12}=\theta_{1}-\theta_{2}$ is the
rapidity difference.
Assuming the commutativity of the supercharges and S-matrix, Zamolodchikov
derived the S-matrix [19].

  The S-matrix can be parametrized as
$$\eqalign{
& |K_{0a}(\theta_{1})K_{a0}(\theta_{2})\rangle_{in} =
A_{0}(\theta_{12})|K_{0a}(\theta_{2})K_{a0}(\theta_{1})\rangle_{out}
+A_{1}(\theta_{12})|K_{0-a}(\theta_{2})K_{-a0}(\theta_{1})\rangle_{out}
 \cr
& |K_{a0}(\theta_{1})K_{0a}(\theta_{2})\rangle_{in}
=B_{0}(\theta_{12})|K_{a0}(\theta_{2})K_{0a}(\theta_{1})\rangle_{out} \cr
& |K_{a0}(\theta_{1})K_{0-a}(\theta_{2})\rangle_{in}
=B_{1}(\theta_{12})|K_{a0}(\theta_{2})K_{0-a}(\theta_{1})\rangle_{out} \cr
} \eqno(5.14)$$
where
$$\eqalign{
&A_{0}(\theta)=e^{C\theta}\cosh(\frac{\theta}{4})S(\theta),\quad
A_{1}(\theta)=-ie^{C\theta}\sinh(\frac{\theta}{4})S(\theta) \cr
&B_{0}(\theta)=\sqrt{2}e^{-C\theta}\cosh(\frac{\theta-i\pi}{4})S(\theta)
, \quad
B_{1}(\theta)=\sqrt{2}e^{-C\theta}\cosh(\frac{\theta+i\pi}{4})S(\theta) \cr
} \eqno(5.15)$$
with
$$\eqalign{
&C=\frac{1}{2\pi i}\log2 \cr
&S(\theta)={1\over \sqrt{\pi}}\prod_{k=1}^{\infty}
\frac{\Gamma(k-\theta/2\pi i)\Gamma(-1/2+k+\theta/2\pi i)}
{\Gamma(1/2+k-\theta/2\pi i)\Gamma(k+\theta/2\pi i)} \cr
}\eqno(5.16)$$

 As for the bosonic sine-Gordon theory, the S-matrix is well known [21].
For kink $A^{+}$ and anti-kink $A^{-}$ we have
$${S_{SG}}_{ab;cd}=\frac{U(\theta)}{i \pi} S_{ab;cd} \eqno(5.17)$$
where $a,b,c,d=\pm$ and ${S_{SG}}_{ab;cd}$ is the scattering amplitude
for the process: $A^{a}+A^{b} \rightarrow A^{c}+A^{d}$ and
$U(\theta)$ is given as
$$\eqalign {
& U(\theta)=\Gamma(\frac{1}{\gamma})
\Gamma(1+i\frac{\theta}{\gamma})\Gamma(1-\frac{1}{\gamma} -
i\frac{\theta}{\gamma}) \prod_{n=1}^{\infty} \frac{R_{n}(\theta)R_{n}
(i \pi -\theta)}{R_{n}(0)R_{n}(i \pi)} \cr
&  R_{n}(\theta)=\frac{\Gamma(\frac{2n}{\gamma} +\frac{i \theta}{\gamma})
  \Gamma(1+\frac{2n}{\gamma} +\frac{i \theta}{\gamma})}
{\Gamma(\frac{2n+1}{\gamma}+ \frac{i \theta}{\gamma})
\Gamma(1+\frac{2n-1}{\gamma}+ \frac{i \theta}{\gamma})}\cr
}\eqno(5.18)$$
with $\gamma$ being given by eq.(3.13).  Non-vanishing $S_{ab;cd}$'s are
$$\eqalign{
&S_{++;++}=S_{--;--}=\sinh [\frac{1}{\gamma}(i \pi-\theta)] \cr
&S_{+-;+-}=S_{-+;-+}=\sinh \frac{\theta}{\gamma} \quad , \quad
S_{+-;-+}=S_{-+;+-}=i \sin \frac{\pi}{\gamma} \cr
}\eqno(5.19)$$
Here we note that the $S_{ab;cd}$ is the well-known solution of Yang-Baxter
equation for the six-vertex model.

  Now we present explict results for minimal N=2 supersymmetric parts of
S-matrices.  First we write down the in-states of tri-critical Ising soliton
and anti-soliton scattering in terms of the out-states. This can be
provided by the matrices ${\cal A}$ and ${\cal B}$ given below. And then,
by taking the tensor product of these in-states
we can construct the S-matrices for soliton-antisoliton scattering for
N=2 supersymmety as follows [22]:
$$\eqalign{
|X{\bar Y}(\theta_1){\bar V}W(\theta_2)\rangle_{in} &=
|X(\theta_1){\bar V}(\theta_2)\rangle_{in}
|{\bar Y}(\theta_1)W(\theta_2)\rangle_{in} \cr
& = \sum_{X',{\bar Y}',{\bar V}', W'}
{\cal B}_{X{\bar V},X'{\bar V}'}(\theta_{12})
{\cal A}_{{\bar Y}W,{\bar Y}'W'}(\theta_{12})
|X'{\bar Y}'(\theta_2){\bar V}'W'(\theta_1)\rangle_{out} \cr
}\eqno(5.20)$$
where the indices $X'$ (${\bar Y}'$) and $W'$ (${\bar V}'$) run over $B$
(${\bar B}$) and $F$ (${\bar F}$). In the above equation the
tri-critical Ising soliton amplitudes ${\cal A}$ and ${\cal B}$ are given by
$$
{\cal A} =
\left(\begin{array}{cccc}
A_{+} & A_{+} & 0 & 0 \\
A_{+} & A_{+} & 0 & 0 \\
0 & 0 & A_{-} & A_{-}  \\
0 & 0 & A_{-} & A_{-}
\end{array}\right)
\hspace{1 cm}
{\cal B} =
\left(\begin{array}{cccc}
B_{+} & B_{-} & 0 & 0 \\
B_{-} & B_{+} & 0 & 0 \\
0 & 0 & B_{+} & B_{-}  \\
0 & 0 & B_{-} & B_{+}
\end{array}\right)
\eqno(5.21)$$
where the rows and columns of ${\cal A}$ are arranged in the order:
${\bar B}B$, ${\bar F}F$, ${\bar B}F$ and${\bar F}B$
, while those of ${\cal B}$ are in the order: $B{\bar B}$, $F{\bar F}$,
$B{\bar F}$ and $F{\bar B}$ . In (5.21)
$A_{\pm}(\theta)$ and $B_{\pm}(\theta)$ are obtained from (5.15) as
$$\eqalign{
& A_{+}(\theta) = \textstyle{1\over 2}(A_{0}(\theta) + A_{1}(\theta))
= \textstyle{1\over 2}e^{C\theta}(\cosh{{\theta\over 4}}
-i \sinh{{\theta\over 4}})S(\theta) \cr
& A_{-}(\theta) = \textstyle{1\over 2}(A_{0}(\theta) - A_{1}(\theta))
= \textstyle{1\over 2}e^{C\theta}(\cosh{{\theta\over 4}}
+i \sinh{{\theta\over 4}})S(\theta) \cr
& B_{+}(\theta) = \textstyle{1\over 2}(B_{0}(\theta) + B_{1}(\theta))
= e^{-C\theta}\cosh{{\theta\over 4}}S(\theta) \cr
& B_{-}(\theta) = \textstyle{1\over 2}(B_{0}(\theta) - B_{1}(\theta))
= - i e^{-C\theta}\sinh{{\theta\over 4}}S(\theta) \cr
}\eqno(5.22)$$
For illustration, we consider the scattering of $B{\bar B}$
and ${\bar B}B$ as follows:
$$\eqalign{
&|B{\bar B}(\theta_1){\bar B}B(\theta_2)\rangle_{in} =
|B(\theta_1){\bar B}(\theta_2)\rangle_{in}
|{\bar B}(\theta_1)B(\theta_2)\rangle_{in} \cr
& = B_{+}A_{+}|(B{\bar B})({\bar B}B)\rangle_{out}
+ B_{-}A_{+}|(F{\bar B})({\bar F}B)\rangle_{out} \cr
& + B_{+}A_{+}|(B{\bar F})({\bar B}F)\rangle_{out}
+ B_{-}A_{+}|(F{\bar F})({\bar F}F)\rangle_{out} \cr
}\eqno(5.23)$$
Hence, for example, we obtain the following result:
$$\eqalign{
& S(B{\bar B} + {\bar B}B \rightarrow B{\bar B} + {\bar B}B) \cr
& = S(B{\bar B} + {\bar B}B \rightarrow B{\bar F} + {\bar B}F) \cr
& =B_{+}A_{+}
=\textstyle{1\over 4}(1+\cosh{\theta\over 2}- i\sinh{\theta\over 2})
S^2(\theta) \cr
}\eqno(5.24)$$
The S-matrices for the processes: ${\bar X}Y +V{\bar W} \rightarrow
{\bar X}'Y' +V'{\bar W}'$ are obtained by crossing symmetry.
The total S-matrices are given as a product of above N=2 supersymmetry parts
wih the ordinary sine-Gordon S-matrix factors as follows:
$$S_{N2SG}(\theta)=S_{MN2}(\theta) \times S_{SG}(x=e^{\theta/\gamma},
q=-e^{-i\pi/\gamma})\eqno(5.25)$$

  Here we have a comment on the correspondence between the coset and the
bosonic sine-Gordon theories.
As we mentioned before, the N=2 supersymmetry is realized for the special
value of the coupling constant, $\beta={2\over \sqrt{3}}$.
By setting $\gamma=2$ corresponding to the above value of $\beta$ in
eqs.(5.17-19)(or setting $\gamma=16\pi$ in the S-matrix formula of ref.[21]
where the normalization of $\gamma$ differs from ours by a factor $8\pi$)
for ordinary bosonic sine-Gordon theory, we obtain the
scattering matrices for soliton $A$ and anti-soliton ${\bar A}$ (here we
use this notation instead of $A^{\pm}$, following Zamolodchikov's notation)
[21]:
$$\eqalign{
&|A(\theta_1){\bar A}(\theta_2)\rangle_{in}
=S_{T}(\theta_{12})|{\bar A}(\theta_2)A(\theta_1)\rangle_{out}
+S_{R}(\theta_{12})|A(\theta_2){\bar A}(\theta_1)\rangle_{out} \cr
& |A(\theta_1)A(\theta_2)\rangle_{in}
=S_{0}(\theta_{12})|A(\theta_2)A(\theta_1)\rangle_{out},
|{\bar A}(\theta_1){\bar A}(\theta_2)\rangle_{in}
=S_{0}(\theta_{12})|{\bar A}(\theta_2){\bar A}(\theta_1)\rangle_{out} \cr
}\eqno(5.26)$$
where $S_T$ and $S_R$ are transmission and reflection amplitudes for soliton-
antisoliton scattering and $S_0$ denotes the amplitude for identical soliton
scattering
$$
 S_{T}(\theta)= -i\sinh{\theta\over 2}S^2(\theta) \quad,
\quad S_{0}(\theta)= \cosh{\theta\over 2}S^2(\theta) \quad,
\quad S_{R}(\theta)= S^2(\theta)
\eqno(5.27)$$
What is remakable here is that we can recover these results by expressing
the soliton states in terms of the coset soliton states as
$$\eqalign{
|A\rangle &= |{\bar B}B\rangle -|{\bar F}B\rangle + |{\bar B}F\rangle +
|{\bar F}F\rangle+|B{\bar B}\rangle + |F{\bar B}\rangle -|B{\bar F}\rangle
+|F{\bar F}\rangle \cr
& = |K_{01}K_{-10}\rangle + |K_{01}K_{10}\rangle + |K_{-10}K_{0-1}\rangle
+|K_{-10}K_{01}\rangle + |K_{0-1}K_{-10}\rangle - |K_{0-1}K_{10}\rangle \cr
& -|K_{10}K_{0-1}\rangle + |K_{10}K_{01}\rangle \cr
}\eqno(5.28)$$
This strongly supports the validity of our calculation for S-matrices of
N=2 SG soliton scattering.

  Now let us briefly consider the N=2 sine-Gordon breathers which are bound
states of the N=2 kinks.  As in the case of N=2 kinks, the N=2 breathers can
be constructed from two tri-critical Ising models and bosonic sine-Gordon
theory. In the case of N=1 sine-Gordon theory we have N=1 bosonic or
fermionic breathers with no topological charge denoted by $\phi_{n}$ and
$\psi_{n}$ which form a supermultiplet of tri-critical Ising model as
discussed by Ahn [20]. The n-th breathers $\phi_{n}$ and $\psi_{n}$ possess
a mass : $m_n = 2m\sin(n\pi\gamma/2)$ ($\gamma <1$) where m is the soliton
mass and set to unity. In our N=2 case, the breathers can be represented as
$$|B_{n}x_{n}y_{n}\rangle \eqno(5.29)$$
where $B_{n}$ is the n-th breather of ordinary sine-Gordon and $x_{n}$ and
$y_{n}$ stand for either $\phi_{n}$ or $\psi_{n}$. Therefore we get the
following N=2 breather multiplet:
$$\eqalign{
&|\Phi^{(00)}_n\rangle \equiv |B_n\phi_n\phi_n\rangle \quad, \quad
|\Phi^{(01)}_n\rangle \equiv |B_n\phi_n\psi_n\rangle \cr
&|\Phi^{(10)}_n\rangle \equiv |B_n\psi_n\phi_n\rangle \quad, \quad
|\Phi^{(11)}_n\rangle \equiv |B_n\psi_n\psi_n\rangle \cr
}\eqno(5.30)$$

  We now examine how to construct S-matrices for breather-soliton scattering.
Let us take $|\Phi^{(00)}_n\rangle$ and $|B{\bar B}\rangle$ as an
initial state.
For minimal N=2 supersymmetric part, we calculate the product of in-states of
$|\phi_n B\rangle$ and $|\phi_n{\bar B}\rangle$, which are given as
$$\eqalign{
&|\phi_n B\rangle_{in} = X_n(\theta)2^{-\Delta\theta_n/2\pi i}
(\alpha_n|\phi_n B\rangle_{out} + \beta_n|\psi_n F\rangle_{out}) \cr
&|\phi_n {\bar B}\rangle_{in} = X_n(\theta)2^{\Delta\theta_n/2\pi i}
(\gamma_n|\phi_n {\bar B}\rangle_{out} +
\delta_n|\psi_n {\bar F}\rangle_{out}) \cr
}\eqno(5.31)$$
where $\Delta\theta_n=i\pi-in\pi\gamma$ and $X_n(\theta)$ is given by
$$
X_n(\theta)=-2S(\theta+{1\over 2}\Delta\theta_n)
S(\theta-{1\over 2}\Delta\theta_n)
\eqno(5.32)$$
with $S(\theta)$ being given in (5.16) and
$\alpha_n=\gamma_n=\cosh{[(2\theta-i\pi)/4]}$, $\beta_n=\sqrt{m_n}$ and
$\delta_n=-i\sqrt{m_n}$ as obtained by Ahn [20].
Hence we get
$$\eqalign{
|\phi_n(\theta_1)B(\theta_2)\rangle_{in}
|\phi_n(\theta_1){\bar B}(\theta_2)\rangle_{in}
&= \bigl (X_n(\theta) \bigr )^2 \bigl \{
\alpha_n\gamma_n|\phi_n\phi_n B{\bar B}\rangle_{out}
+\beta_n\gamma_n|\psi_n\phi_n F{\bar B}\rangle_{out} \cr
&+\alpha_n\delta_n|\phi_n\psi_n B{\bar F}\rangle_{out}
+\beta_n\delta_n|\psi_n\psi_n F{\bar F}\rangle_{out} \bigr \} \cr
}\eqno(5.33)$$
Here, note that the factor $2^{-\Delta\theta_n/2\pi i}$ in $|\phi_n B\rangle_
{in}$ is canceled by the factor $2^{\Delta\theta_n/2\pi i}$ in
$|\phi_n {\bar B}\rangle_{in}$.  Thus we obtain the following S-matrices.
$$\eqalign{
&S(\phi_n\phi_n+B{\bar B} \rightarrow \phi_n\phi_n+B{\bar B})
= \bigl (X_n(\theta) \bigr )^2 \cosh^2{[(2\theta-i\pi)/4]} \cr
&S(\phi_n\phi_n+B{\bar B} \rightarrow \phi_n\psi_n+B{\bar F})
= \bigl (X_n(\theta) \bigr )^2 (-i\sqrt{m_n})\cosh{[(2\theta-i\pi)/4]} \cr
&S(\phi_n\phi_n+B{\bar B} \rightarrow \psi_n\phi_n+F{\bar B})
= \bigl (X_n(\theta) \bigr )^2 \sqrt{m_n}\cosh{[(2\theta-i\pi)/4]} \cr
&S(\phi_n\phi_n+B{\bar B} \rightarrow \psi_n\psi_n+F{\bar F})
= \bigl (X_n(\theta) \bigr )^2 (-im_n) \cr
}\eqno(5.34)$$
Other S-matrices for breather-soliton as well as breather-breather scattering
should also be calculated through a similar procedure.

\vspace{0.8 cm}
\leftline{\large \bf 6. Conclusion}
\vspace{0.8 cm}

  In this article we have investigated the N=2 sine-Gordon theory
in the framework of perturbation theory.
  We obtained the quantum conserved charges which generate $sl(2)$ quantum
affine Kac-Moody algebra $\widehat{sl_{q}(2)}$.  The N=2 supersymmetry
commutes with this quantum group symmetry.

  Based on this two commuting quantum symmetries, we have constructed
S-matrix
of N=2 sine-Gordon theory as a product of S-matrix for the N=2 minimal
supersymmetric part and that for the ordinary sine-Gordon part.  One
interesting point is the topological charges of the two symmetries.  We
derived the relation of the ordinary sine-Gordon topological charge and
the topological charges of N=2 supersymmetry, which was originally discussed
by
Witten and Olive [18].  It turns out that the relation between the
topological
and super-toplogical charges plays an essential role when we restrict the
fundamental particles from the abstract realization theory.
One remarkable observation is that we can reproduce the soliton-antisoliton
S-matrices of ordinary bosonic sine-Gordon theory at a special value of the
coupling constant, $\beta={2\over \sqrt3}$, which corresponds to the N=2 SUSY
point.

   Now some remarks on further problems are in order.
Our model is the simplest model and we believe it can be extended to any N=2
supersymmetric models [23-25]. In this regards, it would be intriguing to
extend the present analysis to the N=2 supersymmetric version of Toda field
theories [26-31], to see if the quantum group structure is described by
commuting N=2 supersymmetry and quantum group symmetry $\widehat{sl_{q}(n)}$
for $n=3,4,\cdots$. Another interesting subject is to investigate the
correspondence of N=2 sine-Gordon theory with some Thirring-type fermionic
theories.  This is motivated by the well-known correspondence between the
N=0 sine-Gordon theory and the massive Thirring model [32,33].
We have one more comment on supersymmetry (see refs.[13,17,23,24]).
We think it becomes
clear that S-matrix of N-extended supersymmetric models can be constructed
simply by
making the N product of that of the N=1 supersymmetric model (tri-critical
Ising model) and some bosonic part, up to CDD ambiguity.  We are interested
in whether there is a model for arbitrary N.

\bigskip

We thank T.~Eguchi, T.~Inami, A.~LeClair and S.~K.~Yang for useful
discussions.
One of us (K.~K.) would like to thank the members of high-energy theory group
at Purdue University and the theory group in College of Liberal Arts and
Sciences, Kyoto University for their kind hospitality.
This work was partially supported by the Grant-in-Aid for Scientific Research
from the Ministry of Education, Science and Culture (\#63540216).

\newpage
\vspace{0.8 cm}

\vspace{0.8 cm}
\centerline{\large \bf  References}
\vspace{0.8 cm}

\begin{description}

\item{[1]}
A.~B.~Zamolodchikov, Sov.~J.~Nucl.~Phys.~{\bf 44}(1986)529; Reviews in
\hfill\break
Mathematical Physics {\bf 1}(1990)197.

\item{[2]}
T.~Eguchi and S.~K.~Yang, Phys.~Lett.~{\bf 224B}(1989)373 and {\bf 235B}
(1990)282.

\item{[3]}
F.~A.~Smirnov, Int.~J.~Mod.~Phys.~{\bf A4}(1989)4213; Nucl.~Phys.~{\bf B337}
(1990)156; \hfill\break
N.~Reshetikhin and F.~Smirnov, Commun.~Math.~Phys.~{\bf 131}(1990)157.

\item{[4]}
A.~LeClair, Phys.~Lett.~{\bf 230B}(1989)103.

\item{[5]}
V.~A.~Fateev and A.~B.~Zamolodchikov, Int.~J.~Mod.~Phys.~{\bf A6}(1990)1025.

\item{[6]}
D.~Bernard and A.~LeClair, Nucl.~Phys.~{\bf B340}(1990)721; Phys.~Lett.
\hfill\break {\bf 247B}(1990)309.

\item{[7]}
R.~Sasaki and I.~Yamanaka, in Advanced Studies in Pure Mathematics {\bf 16}
(1988)271.
; Prog.~Theor.~Phys. {\bf 79}(1988)1167.

\item{[8]}
V.~G.Drinfel'd, Sov.~Math.~Dokl.~{\bf 32}(1985)254.

\item{[9]}
M.~Jimbo, ~Lett.~Math.~Phys.~{\bf 10}(1985)63.

\item{[10]}
D.~Bernard and A.~LeClair, Commun.~Math.~Phys.~{\bf 142}(1991)99.

\item{[11]}
P.~Mathieu, Nucl.~Phys.~{\bf B336}(1990)338.

\item{[12]}
C.~Ahn, D.~Bernard and A.~LeClair, Nucl.~Phys.~{\bf B346}(1990)409.

\item{[13]}
K.~Schoutens, Nucl.~Phys.~{\bf B344}(1990)665.

\item{[14]}
K.~Kobayashi and T.~Uematsu, Phys.~Lett.~{\bf 264B}(1991)107.

\item{[15]}
K.~Kobayashi, T.~Uematsu and Y.-Z.~Yu, \lq\lq Quantum Conserved Charges
in N=1 and N=2 Supersymmetric Sine-Gordon Theories \rq\rq preprint KUCP-29/91;
 PURD-TH-91-03 (1991).

\item{[16]}
T.~Inami and H.~Kanno, Commun.~Math.~Phys.~{\bf 136}(1991)519; Nucl.~Phys.
\hfill\break {\bf B359}(1991)201.

\item{[17]}
R.~Shankar and E.~Witten, Phys.~Rev.~{\bf D17}(1978)2134; Nucl.~Phys.
\hfill\break {\bf B141}(1978)525.

\item{[18]}
E.~Witten and D.~Olive, Phys.~Lett.~{\bf 78B}(1978)97.

\item{[19]}
A.~B.~Zamolodchikov, \lq\lq Fractional-Spin Integrals of Motion in Perturbed
Conformal Field Theory \rq\rq , Moscow preprint (1989).

\item{[20]}
C.~Ahn, Nucl.~Phys.~{\bf B354}(1991)57.

\item{[21]}
A.~B.~Zamolodchikov and A.~B.~Zamolodchikov, Ann.~Phys.~(NY)~{\bf 120}
\hfill\break (1979)253.

\item{[22]}
K.~Kobayashi and T.~Uematsu, \lq\lq S-matrix of N=2 Supersymmetric Sine-Gordon
Theory \rq\rq, preprint KUCP-38;PURD-TH-91-09 (1991),
Phys.~Lett.~{\bf B} to be published.

\item{[23]}
P.~Fendley, S.~D.~Mathur, C.~Vafa and N.~P.~Warner,
Phys.~Lett.~{\bf 243B}(1990)257.

\item{[24]}
P.~Fendley, W.~Lerche, S.~D.~Mathur and N.~P.~Warner, Nucl.~Phys.~{\bf B348}
(1991)66.

\item{[25]}
P.~Mathieu and M.~A.~Walton, \lq\lq Integrable Perturbations of N=2
Superconformal Minimal Models \rq\rq , Quebec preprint(1990).

\item{[26]}
T.~J.~Hollowood and P.~Mansfield, Phys.~Lett.~{\bf 226B}(1989)73.

\item{[27]}
J.~Evans and T.~Hollowood, Nucl.~Phys.~{\bf B352}(1991)723.

\item{[28]}
H.~W.~Braden, E.~Corrigan, P.~E.~Dorey and R.~Sasaki, Nucl.~Phys.~{\bf B338}
(1990)689.

\item{[29]}
T.~Nakatsu, Nucl.~Phys.~{\bf B356}(1991)499.

\item{[30]}
G.~W.~Delius, M.~T.~Grisaru, S.~Penati and D.~Zanon, Nucl.~Phys.~{\bf B359}
(1991)125.

\item{[31]}
S.~Komata, K.~Mohri and H.~Nohara, Nucl.~Phys.~{\bf B359}(1991)168.

\item{[32]}
S.~Coleman, Phys.~Rev.~{\bf D11}(1975)2088.

\item{[33]}
S.~Mandelstam, Phys.~Rev.~{\bf D11}(1975)3026.

\end{description}

\end{document}